\documentclass[12pt,twoside]{article}
\usepackage{graphicx,epsf,epsfig,rotating}

\begin{document}

\newcommand{\lsim}   {\mathrel{\mathop{\kern 0pt \rlap
  {\raise.2ex\hbox{$<$}}}
  \lower.9ex\hbox{\kern-.190em $\sim$}}}
\newcommand{\gsim}   {\mathrel{\mathop{\kern 0pt \rlap
  {\raise.2ex\hbox{$>$}}}
  \lower.9ex\hbox{\kern-.190em $\sim$}}}
\def\be{\begin{equation}}
\def\ee{\end{equation}}
\def\ba{\begin{eqnarray}}
\def\ea{\end{eqnarray}}
\def\d{{\rm d}}
\def\ap{\approx}
\def\for{\qquad {\rm for} \qquad}
\def\t{\vartheta}
\def\eps{\varepsilon}

\sloppy

\title{\hfill {\small MPI--PhT 2003-27}\\
       \vskip1.0cm
        Superheavy dark matter as UHECR source\\ versus the SUGAR data}

\author{M.~Kachelrie{\ss}$^{1}$ and D.~V.~Semikoz$^{1,2}$ \\
        {\it\small $^1$Max-Planck-Institut f\"ur Physik
        (Werner-Heisenberg-Institut),}\\
         {\it\small F\"ohringer Ring 6, 80805 M\"unchen, Germany} \\
        {\it\small $^2$Institute for Nuclear Research of the Academy
         of Sciences of Russia, Moscow, 117312, Russia}  }
 
\date{}

\maketitle
\begin{abstract}
Decay or annihilation products of superheavy dark matter (SHDM) 
could be responsible for the end of the Ultra-High Energy
Cosmic Ray (UHECR) spectrum. In this case, the south array of the
Pierre Auger Observatory should observe in the future a significant
anisotropy of UHECR arrival directions towards the galactic
center. Here we use the already existing data of the SUGAR array
to test this possibility. 
If decaying SHDM is distributed according a Navarro-Frenk-White (NFW)
dark matter profile with core radius $R_c=15$ kpc 
and is responsible only for UHECRs above $\sim
6\times 10^{19}$~eV, i.e. the AGASA excess, then the arrival
directions measured by the SUGAR array have a probability of
$\sim 10\%$ to be consistent with this model. By contrast, the model of
annihilating SHDM is disfavoured at least at 99\% CL by the SUGAR
data, if the smooth component of the DM dominates the signal. 
\end{abstract}

{PACS numbers: 98.70.Sa, 14.80.-j}   

\section{Introduction}
Protons accelerated by uniformly distributed extragalactic 
astrophysical sources would be a perfect minimal explanation  of the
UHECR data above $10^{19}$~eV. However, protons with energy $E >
4\times 10^{19}$~eV loose quickly energy due to pion production on
cosmic microwave background photons. Thus the proton spectrum should
show the so-called Greisen-Zatsepin-Kuzmin (GZK) cutoff~\cite{gzk},
which is not observed by the AGASA experiment~\cite{agasa}.  
Moreover, if the small-scale clusters in the arrival directions of
UHECR measured by AGASA~\cite{AGASAcluster_data} are due to point-like
sources, one can estimate their number~\cite{Dubovsky2000}.
This number is so small that the nearest source should be located at
the distance $R_{\rm min} \sim 100$~Mpc~\cite{kst2003}. This means
that the GZK cutoff is exponentially sharp at $E\approx 6\times 10^{19}$~eV
and even the data of the HiRes experiment~\cite{Hires} are inconsistent
with the expected proton spectrum~\cite{kst2003}.  This inconsistency
becomes even stronger if BL Lacs, which show a statistically
significant correlations with the arrival directions of UHECR with energy
$E \sim 4-6\times 10^{19}$~eV \cite{corr_bllac}, are sources of
UHECR. 

A possible solution to this problem would be the existence of
superheavy dark matter (SHDM)~\cite{bkv97,kr97}. Superheavy particles
with mass $M_X \sim 10^{13-14}$ GeV can be naturally produced during
inflation and would be today the dominant component of dark
matter~\cite{kt99}. Such particles will concentrate 
in galactic halos and the secondaries from their decay could be
responsible for the highest energy cosmic rays. It has also been
suggested that not decays but annihilations of SHDM particles produce
the observed UHECRs~\cite{BDK}.   

By construction, this model has two clean signatures: the dominance
of photons at the highest energies~\cite{bkv97} and an anisotropy of the
arrival directions with an increased flux from the Galactic
center~\cite{dt98,bbv98}. Unfortunately, both signatures are not very
clean for the present experiments. Indeed, at 95\% C.L., $\sim 30\%$
of the UHECR above $E>10^{19}$~eV can be photons~\cite{photon_limit},
which means that still most of UHECRs with $E > 4\times 10^{19}$~eV 
can be photons without any contradiction to the experimental
data. Since experiments in the northern hemisphere do not see the
Galactic center, they are not very  sensitive to a possible anisotropy
of arrival directions of UHECR from SHDM.  In contrast,  the Galactic
center was visible for the old Australian SUGAR experiment.

The anisotropy of the arrival directions using data from the full sky
was discussed in 
Refs.~\cite{an1,an2}. Reference~\cite{an1} compared the flux from the
Galactic center to the one from the anti-center and found them to be
comparable. Similarly, the full-sky harmonic analysis including AGASA and
SUGAR data from Ref.~\cite{an2} found no significant anisotropy.
In this work, we use a two-component energy spectrum    
of UHECRs consisting of protons from uniformly distributed,
astrophysical sources and the fragmentation products of SHDM
calculated in SUSY-QCD. We compare their expected arrival direction
distribution to the data of the SUGAR experiment using a
Kolmogorov-Smirnov test. Contrary to the harmonic analysis, this test
allows to quantify directly the (dis-) agreement of the measured
distribution of arrival direction with the expected one in the SHDM
model. We consider both decays and annihilations
of SHDM.

The paper is organized as follows: In Section II we discuss the status
of the SUGAR data. The contribution of SHDM to the UHECR spectrum is
discussed in Section III. In Section IV we perform an harmonic
analysis of the arrival directions measured by the SUGAR experiment.
Then we use a Kolmogorov-Smirnov test in one and two
dimensions to check the consistency of the SUGAR data with the
probability distribution of arrival directions expected for SHDM
in Section V. Finally we conclude in Section VI.

\section{Assessment of the SUGAR data}

\begin{figure}[ht]
\includegraphics[height=0.8\textwidth,clip=true,angle=270]{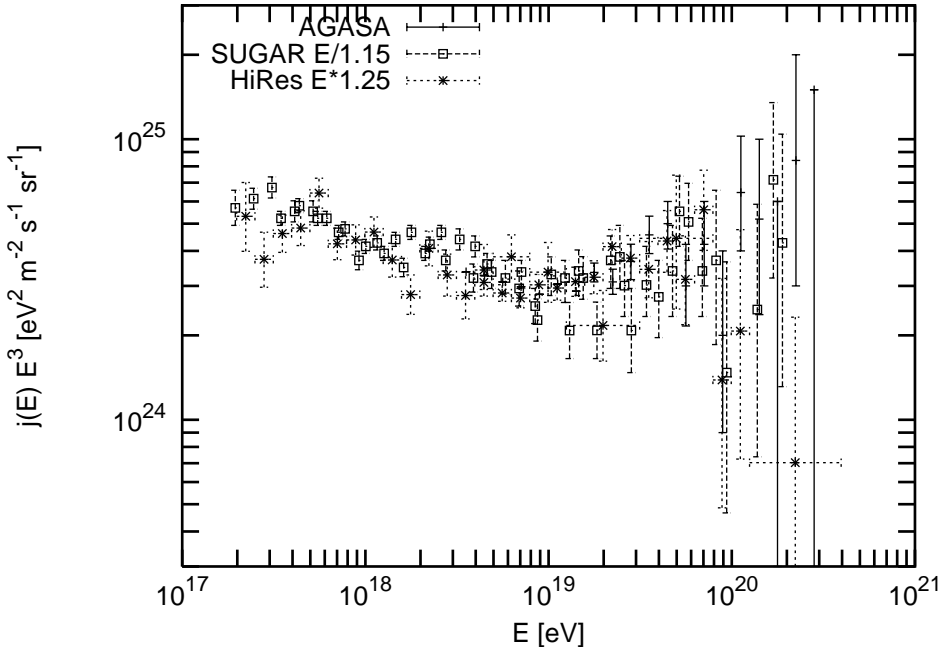}
\caption[...]{UHECR spectrum measured by the AGASA, HiRes and SUGAR
  experiments. We scale the SUGAR spectrum down by $E/1.15$ and the
  HiRes spectrum up by $1.25\times E$ using the AGASA spectrum as
  reference. The overall normalization of the spectra has only a weak
  influence on our results.   
\label{F01}}
\end{figure}

In order to use the SUGAR energy spectrum correctly we compare their
data given in Ref.~\cite{Winn:un} to the energy spectrum measured by
the AGASA~\cite{agasa} and HiRes experiments~\cite{Hires} in
Fig.~\ref{F01}. Rescaling the SUGAR energies calculated with the
Hillas prescription by 15\% downwards, $E_{\rm Hillas} \rightarrow
E_{\rm Hillas}/1.15$, makes their data consistent with the ones from
AGASA. The same is true for the HiRes spectrum if the energy is
rescaled up-wards by 25\%~\footnote{The aperture of HiRes is energy
dependent; the rescaling we perform should be seen therefore just as a
crude approximation.}
We have chosen arbitrary the AGASA spectrum as reference, but changing
the overall normalization of the spectra affects our results only weakly.
As seen from Fig.~\ref{F01}, the SUGAR spectrum
has the ankle at the correct place around $E\approx 10^{19}$~eV and is
consistent with the AGASA spectrum in the whole energy range. In
particular, the SUGAR spectrum also does not show the GZK cutoff     
at the highest energies.

The rescaling of the SUGAR data downwards by 15\% should be compared
to the recent reevaluation of the energy conversion formula used in the
Haverah Park experiment~\cite{ave}. In this reference, the relation
between $\rho(600)$ and the primary energy  has been recalculated using
QGSJET~\cite{QGS} and compared to the original relation suggested by
Hillas. The new calibration results in $\sim 30\%$ lower primary
energies. 

The SUGAR experiment was a very sparse array of detectors and its
energy determination of each single event was therefore rather
unprecise. Thus, we shall use as a statistical test later on a method    
which relies only on the total flux measured by the SUGAR array, but uses
not the energy of each single event.
Since after the rescaling of the energies measured by SUGAR, 
$E_{\rm Hillas} \rightarrow E_{\rm Hillas}/1.15$, its measured flux
is consistent with newer experiments like AGASA and HiRes, we conclude
that {\em on average\/} the energy determination in the SUGAR 
experiment was reliable.
  
The energy conversion formula used in SUGAR to connect the measured
muon number $N_\mu$ with the primary energy assumes that the primary 
is a hadron. For photon primaries, predicted to be dominant in the SHDM
model, the muon content of the shower is smaller by a factor 5 --
10~\cite{muon}. Thus the energies of photon events is expected to be
{\em underestimated\/} by the SUGAR experiment.  The SUGAR spectrum
shown in Fig.~1 would be unchanged at energies 
$E\lsim 5\times 10^{19}$~eV, i.e at energies where all three
experiments agree after rescaling. 

The angular acceptance $\eta(\delta)$ as function of declination
$\delta$ averaged over time of an experiment 
at geographical latitude $b$ ($b=-30.5^\circ$ for SUGAR) observing
showers with maximal zenith angle $\theta_{\max}$ is   
\be  \label{cosz}
 \eta(\delta) \propto \int_0^{\alpha_{\rm \max}} d\alpha\cos(\theta) 
              \propto \left[ \cos(b)\cos(\delta)\sin(\alpha_{\rm \max}) +
                      \alpha_{\rm \max}\sin(b)\sin(\delta) \right]
\ee
where 
\be
 \xi = \frac{\cos(\theta_{\max})-\sin(b) \sin(\delta)} 
            {\cos(b) \cos(\delta)}
\ee
and
\be
 \alpha_{\rm \max} = \left\{ \begin{array}{l}
             \arccos(\xi)  \for -1\leq\xi\leq 1 \,,\\
             \pi           \hspace{1.4cm}\for \xi< -1 \,,\\
              0            \hspace{1.4cm} \for \xi> 1 \,.
             \end{array} \right. 
\ee
We have checked that the zenith angle distribution of the SUGAR events
agrees with the theoretical predicted one, $dN_{\rm th}\propto
d\theta\sin(\theta)\cos(\theta)$, above $E\gsim 4\times 10^{19}$~eV.
At lower energies, the acceptance of the experiment becomes energy
dependent and deviations from $dN_{\rm th}$ start to grow.

\section{Superheavy dark matter contribution to UHECR spectrum}
\label{sec:darkmatter}

We fix the contribution of SHDM to the total UHECR flux following the
assumptions of Ref.~\cite{BGG}: we assume that no galactic
astrophysical sources contribute to the cosmic ray flux above
$10^{19}$~eV and that the extragalactic cosmic ray flux can be
characterized by an injection spectra of protons with a single power
law, $j_{\rm ex}(E)\propto E^{-\alpha}$.  For the choice of
$\alpha=2.7$, this energy spectra modified by redshift, $e^+ e^-$ and
pion production fits very well the measured spectra below $E<6-8\times
10^{18}$~eV~\cite{BGG}. The only difference with \cite{BGG} is that we
take into account that total number of sources is small~\cite{kst2003}, 
if the small scale clusters measured by the AGASA experiment are due
to point-like sources. The AGASA data favor as minimal distance 
to the nearest source $D_{\rm min} \sim 100$~Mpc \cite{kst2003}. Therefore,
the contribution of protons from extragalactic sources has a sharp
cutoff. For the calculations of the proton spectrum we used the
code~\cite{code2}. We use then the SUSY QCD fragmentation functions
$D(x,M_x)$ of superheavy particles with mass $M_X$ calculated in
Ref.~\cite{FF} to model the flux $j_{\rm DM}(E)\propto
D_\gamma(x,M_X)$.  The total UHECR flux is thus
\be 
 j(E)= (1-\epsilon) j_{\rm ex}(E) + \epsilon j_{\rm DM}(E) \,.
\label{spec_norm}
\ee

\begin{figure}[ht]
\includegraphics[height=0.5\textwidth,clip=true,angle=270]{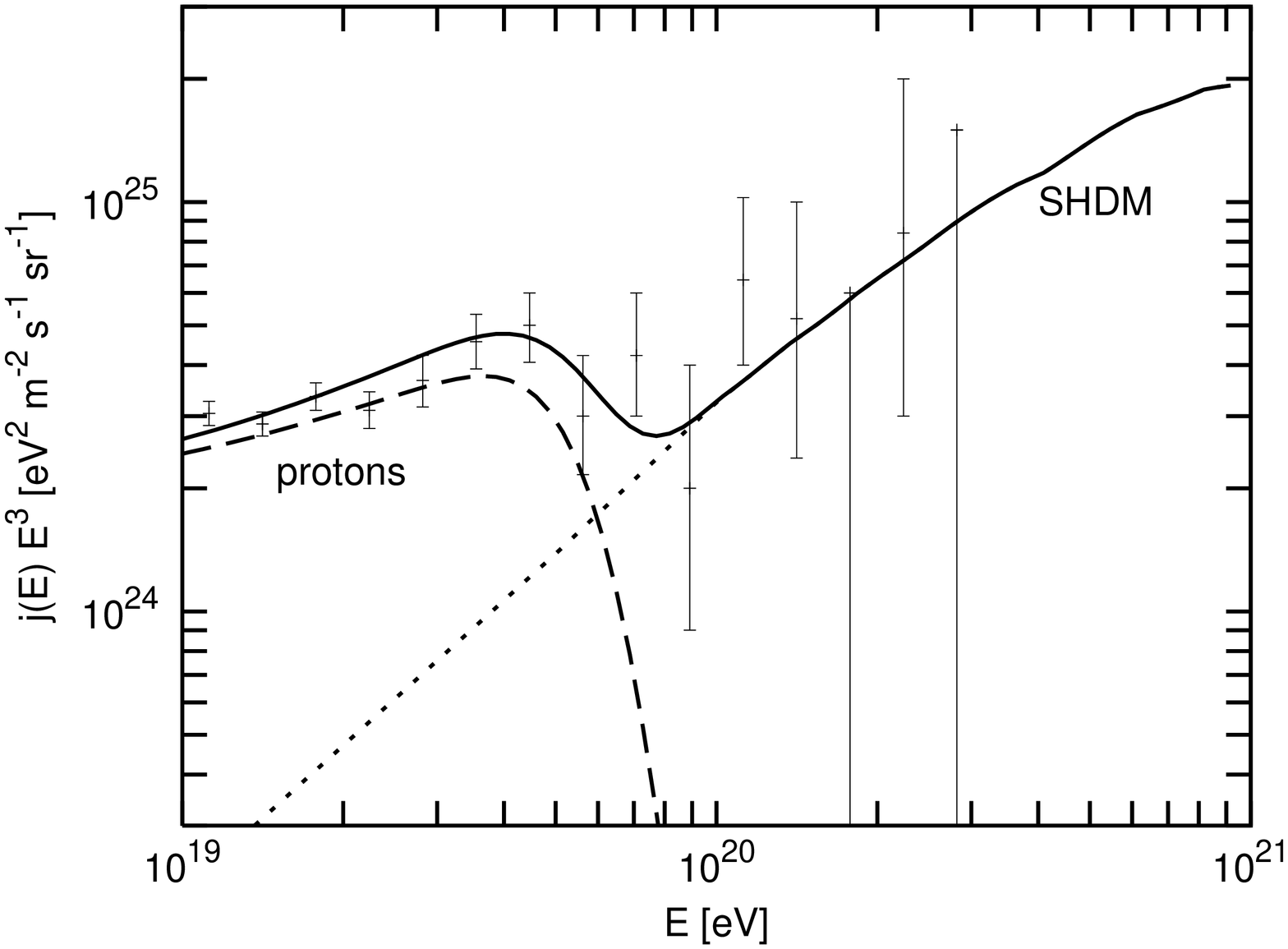}
\includegraphics[height=0.5\textwidth,clip=true,angle=270]{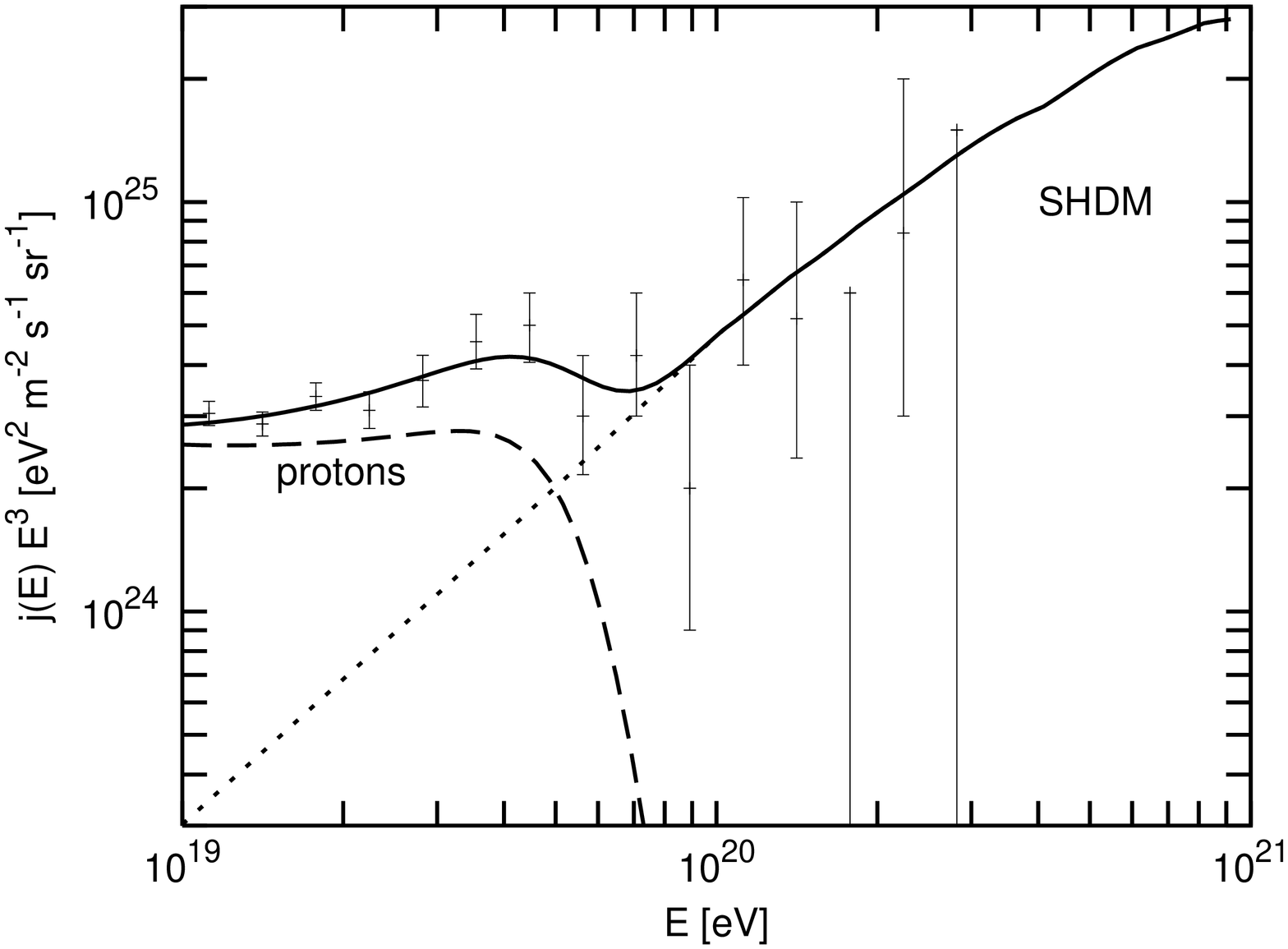}
\caption[...]{UHECR spectrum measured by AGASA experiment. Protons
  from extragalactic sources contribute below the GZK cutoff, photons from
  SHDM decays mainly above the GZK cutoff; for
  hard, $1/E^{2.3}$, (left panel) and soft, $1/E^{2.7}$ (right panel)
  injection spectrum of extragalactic protons. 
\label{F00}}
\end{figure}

We fix the constant $\epsilon$ determining the relative contribution of
SHDM to the UHECR flux by a fit of $j(E)$ to the AGASA
data~\cite{agasa}. In Fig.~\ref{F00}, we show our fits for the case of
a harder $1/E^{2.3}$ (left) and a softer $1/E^{2.7}$ (right) injection
spectrum of extragalactic protons. In the first case, the contribution
of SHDM to the UHECR spectrum below the GZK cutoff is minimal and
starts to dominate only at highest energies $E > 6\times 10^{19}$~eV. 
For this choice of injection spectrum, the contribution of galactic
sources dominate for $E<10^{19}$~eV. In the second case, SHDM gives a
larger contribution at lower energies $E < 6\times 10^{19}$ than
before, and again starts to dominate at $E > 6\times 10^{19}$ eV. 
Note, that most UHECRs above $E>6\times10^{19}$~eV should be photons
in this model, but this does not contradict the rather weak existing
bound of 30\% of photons at $E > 10^{19}$~eV \cite{photon_limit}. 

In the following, we shall use conservatively the case of the
harder $1/E^{2.3}$ injection spectrum if not otherwise stated.
Then the contribution of SHDM to the UHECR spectrum is fixed by
Eq.~(\ref{spec_norm}).

\section{Harmonic analysis} 

In order to compare  SUGAR data with an uniform distribution typical
for extragalactic sources we have performed an one-dimensional
harmonic analysis. As usual we sum  
\be
 a_k = \frac{2}{n} \sum_{a=1}^n \cos(k\phi) \qquad\mbox{and}\qquad
 b_k = \frac{2}{n} \sum_{a=1}^n \sin(k\phi)
\ee
over the $n$ data points.

The amplitude $r_k$ and and phase $\phi_k$ of the $k$.th harmonic are
given by 
\be
 r_k = \sqrt{a_k^2+b_k^2} \qquad\mbox{and}\qquad
 \phi_k=\arctan(b_k/a_k) 
\ee
with chance probability
\be
 p_{\rm ch}=\exp\left( -nr_k^2/4 \right) \,.
\ee
The direction to the signal is $\phi=k\phi_k$.

Results of a harmonic analysis in right ascension $\alpha$ depending on 
$E_{\rm min}$ and  $\theta_{\rm max}$ are given in Table \ref{tab}. 
The results for all harmonics show generally good agreement with an
isotropic distribution for any cutoff energy we have used. Only the
third harmonics shows some anisotropy, in particular at the highest
energies, $E>8\times 10^{19}$~eV; however its phase does not points
towards the galactic center (lying at $\alpha=266^\circ$). Generally,
all harmonics point instead towards $\alpha\sim 130^\circ$. 
Reference~\cite{linsley} derived the probability distribution (pdf) that
a data set with phase $\phi_1$ and amplitude $r_1$ is drawn from an
arbitrary pdf. However, we are not aware of a generalization to
higher harmonics and, in particular, of a method to combine the
information content of several harmonics.
In the next section, we use therefore a Kolmogorov-Smirnov test to
quantify the (dis-) agreement between the expected distribution of
arrival direction and the SUGAR data.

\begin{table}
\begin{center}
\begin{tabular}{|c|c|c||c|c|c||c|c|c|c|} 
\hline 
 $k$  & $\phi$/degree & $p_{\rm ch}$/\% &
 $k$  & $\phi$/degree & $p_{\rm ch}$/\% &
 $k$  & $\phi$/degree & $p_{\rm ch}$/\% \\[0.5ex] \hline
 1 & 111 & 91  & 1 & 157 & 80 & 1 &  124 & 52  \\ \hline
 2 & 130 & 32  & 2 & 117 & 13 & 2 &  140 & 16   \\ \hline
 3 & 131 & 17  & 3 & 124 & 16 & 3 &  119 & 3     \\ \hline
 4 & 151 & 58  & 4 & 133 & 27 & 4 &  136 & 33   \\ \hline
\multicolumn{3}{|c||}{$E_{\rm min} = 4.0\times 10^{19}$~eV} &
\multicolumn{3}{|c|}{$E_{\rm min} = 6.0\times 10^{19}$~eV}  &
\multicolumn{3}{|c|}{$E_{\rm min} = 8.0\times 10^{19}$~eV}  \\ \hline
\multicolumn{3}{|c||}{$\theta_{\rm max} = 55^\circ$} &
\multicolumn{3}{|c||}{$\theta_{\rm max} = 55^\circ$} &
\multicolumn{3}{|c|}{$\theta_{\rm max} = 55^\circ$} \\ \hline
\end{tabular}
\end{center}
\caption{\label{tab} 
Direction $\phi$ to the signal in right ascension and chance
probability of the $k.$th 
harmonics to arise from an isotropic distribution; for different cuts
in energy $E$ and zenith angle $\theta<55^\circ$.}
\end{table}

\section{Kolmogorov-Smirnov tests}
The pdf to detect an event with
energy $E$ and arrival direction $\alpha,\delta$ is a combination of
the isotropic extragalactic and the SHDM flux,
\be   \label{p_alpha3d}
 p(E,\alpha,\delta) \propto \eta(\delta)
 \left[ j_{\rm ex}(E) + 
        j_{DM}(E)\int_0^{s_{\max}} ds \:n_{\rm DM}(r(\alpha,\delta))
 \right]
\ee
where $s_{\max}=R_E\cos\theta+\sqrt{R_h^2-R_E^2\sin^2\theta}$ is given
by the extension $R_h=100$~kpc of the DM halo and $\theta$ is the angle
relative to the direction to the GC.
As explained, the relative size of the two contributions is fixed by
the fit to Eq.~(\ref{spec_norm}).

For the two-dimensional test, we have integrated Eq.~(\ref{p_alpha3d})
over energy,
\be   \label{p_alpha2d}
 P_{2d}(\alpha,\delta) = \int_{E_{\rm min}}^{E_{\rm max}} dE ~
 p(E,\alpha,\delta) \,, 
\ee
where $E_{\rm min}$ and $E_{\rm max}$ are the minimal and maximal
energy considered in the UHECR spectrum. We have used 
$E_{\rm max}=10^{21-22}$ eV, but the results do depend on very weakly
on the exact value. By contrast, the value of $E_{\rm min}$ has a
strong influence on the results obtained.

For the one-dimensional test, we have integrated Eq.~(\ref{p_alpha2d})
over the declination,
\be   \label{p_alpha1d}
 P_{\rm CR}(\alpha) = \int_{-\pi/2}^{\pi/2} d\delta \cos\delta ~
 P_{2d}(\alpha,\delta) \,.
\ee

In the standard one-dimensional Kolmogorov-Smirnov (KS) test, the
maximal difference $D$ between the cumulative probability distribution
function $P(x)=\int dx' p(x')$ and the cumulative distribution
function of the data,  
\be
 S(x) = \frac{1}{n} \sum_i \theta(x_i-x) \,,
\ee
is used as estimator for the belief that the data are drawn from the
distribution $p$. A variant of this test which is equally sensitive on
differences for all $x$ and is especially well suited for data on
$S^1$ uses instead of $D$ the symmetric estimator
\be
 V = D_+ + D_- = \max[S(x)-P(x)] + \max[P(x)-S(x)] \,.
\ee
The significance of a certain value of $V$ is calculated with the
formula given in \cite{numrec}. Since the exposure of a ground-array
experiment is uniform in right ascension $\alpha$, we use $\alpha$ as
variable in the one-dimensional KS test. More exactly, we use as pdf
Eq.~(\ref{p_alpha3d}) integrated over $dE$ and $d\delta \cos\delta$.

\begin{figure}[ht]
\begin{center}
\includegraphics[height=0.45\textwidth,clip=true,angle=270]{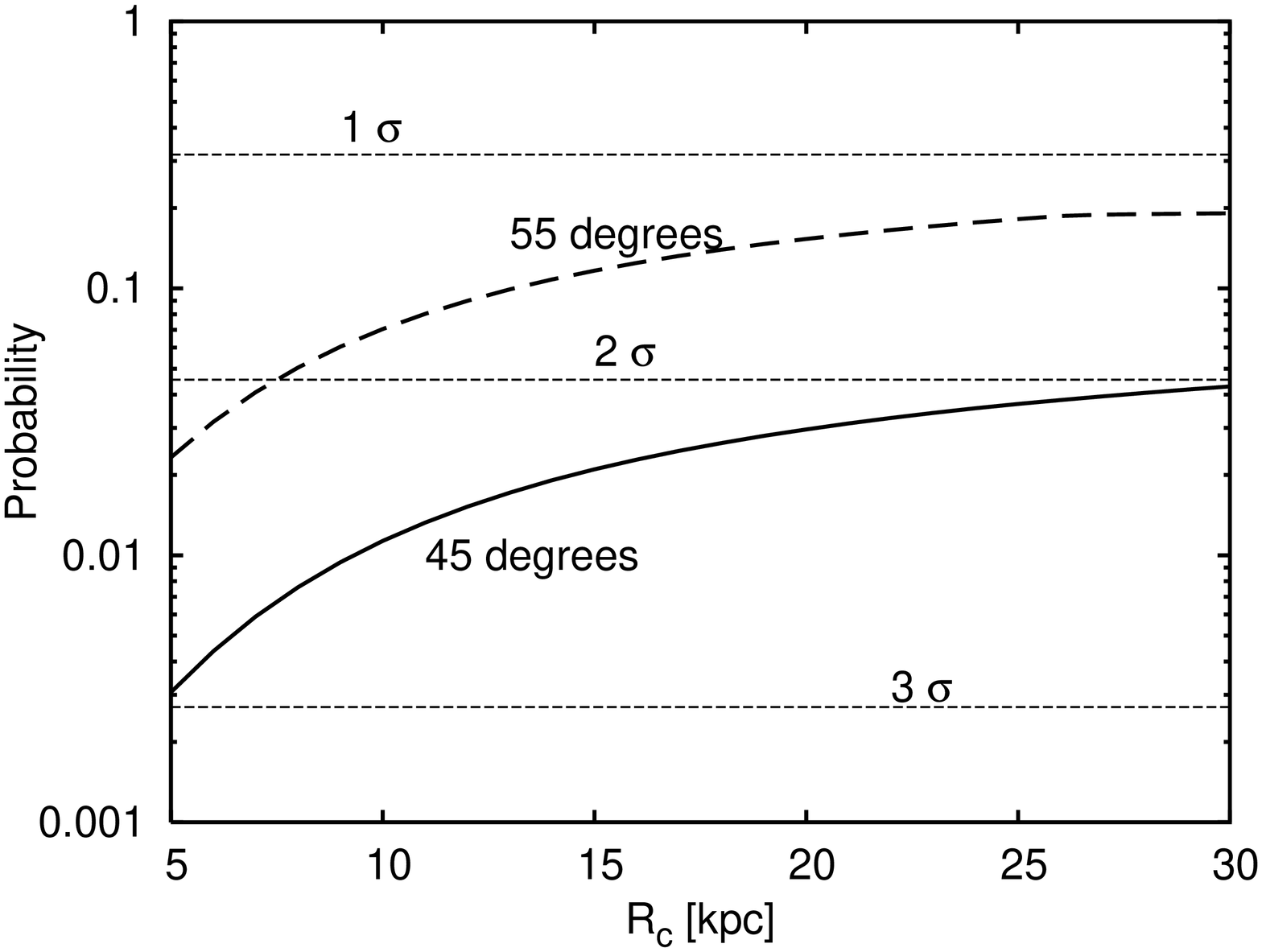}
\includegraphics[height=0.45\textwidth,clip=true,angle=270]{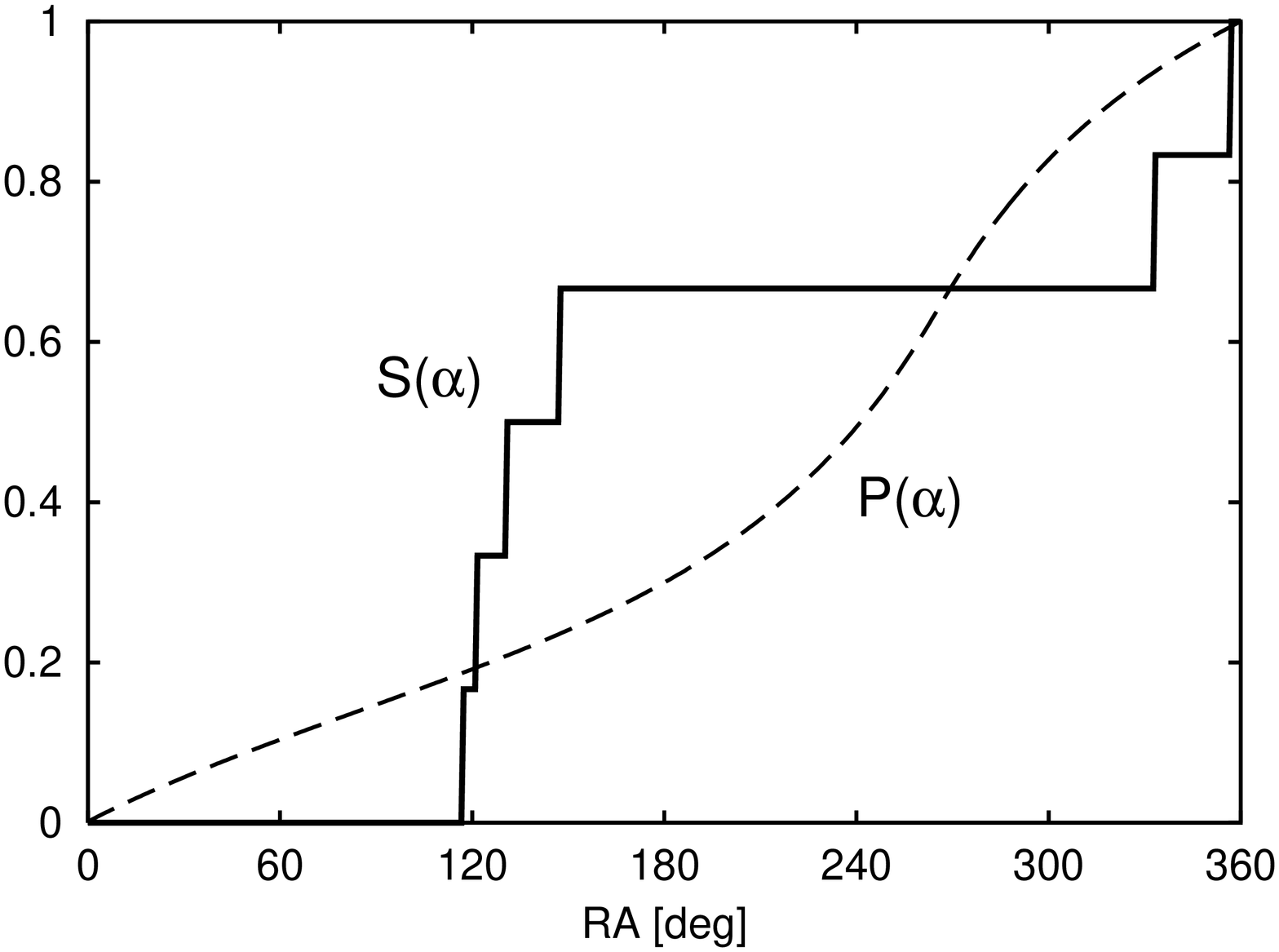}
\caption[...]{Left: 
  Consistency level of the SUGAR data with SHDM distributed
  according a NFW profile as function of the core radius $R_c$; SHDM
  is assumed to be the source of all UHECRs above $E_{\rm
  min}=8\times10^{19}$~eV.  
  Right: Comparison of $S(\alpha)$ and $P(\alpha)$ for $R_s = 15$~kpc,
  $E_{\rm min}=8\times10^{19}$~eV and $\theta_{\rm max}=45^\circ$.
\label{F02}}
\end{center}
\end{figure}

As simplest test, we assume that all SUGAR events above $E_{\rm min}$
are produced by SHDM. Thus we compare $S(\alpha)$ with 
$P(\alpha)=\int_0^\alpha d\alpha'  P_{\rm CR}(\alpha')$. 
The result is shown for $E> 8 \times 10^{19}$~eV in Fig.~\ref{F02}a for
two different values of the  maximal zenith angle, 
$\theta_{\rm max}=45^\circ$ and $\theta_{\rm max}=55^\circ$.
While for $\theta_{\rm max}=45^\circ$ SHDM is disfavoured at
the two sigma level for realistic values of the core radius, $R_c\sim
20$~kpc, the SUGAR data have for the choice of $\theta_{\rm max}=55^\circ$ a
rather large probability $p$ to be consistent with the SHDM
hypothesis, $p\sim 20\%$. 
In Fig.~\ref{F02}b, we compare the two commulative distributions
$S(\alpha)$ and $P(\alpha)$ for $R_s = 15$~kpc, 
$E_{\rm min}=8\times10^{19}$~eV and $\theta_{\rm max}=45^\circ$. 
Inspecting $S(\alpha)$ makes it clear that the data in this case are
not uniformly distributed but clustered around $\alpha\sim 130^\circ$
and $\alpha\sim 350^\circ$. Since none of these two directions
coincide with the position of the GC, this data set disfavours the
SHDM hypothesis more strongly than one would expect for uniformly
distributed events from extragalactic sources. However, one should use
a rather low value of $E_{\rm min}$ to  minimize the uncertainties in
the SUGAR energy determination and we will therefore not rely
on these results.

We consider therefore next as more realistic test the case that both
SHDM and extragalactic sources contribute to the UHECR spectrum. Then 
the dependence of $p$ on $E_{\rm min}$ should be diminished. More
exactly, one would expect in the case that the SHDM hypothesis is
disfavoured by the data that decreasing $E_{\rm min}$ first decreases
$p$. This decrease of $p$ should continue down until $E\sim (3-4)\times
10^{19}$~eV, i.e. until a point where the signal-to-background ratio
becomes considerably smaller than one. 
Decreasing $E_{\rm min}$ even further  should result
in a increase of $p$ because now practically all new events are from
extragalactic sources.

\begin{figure}[ht]
\begin{center}
\includegraphics[height=0.45\textwidth,clip=true,angle=270]{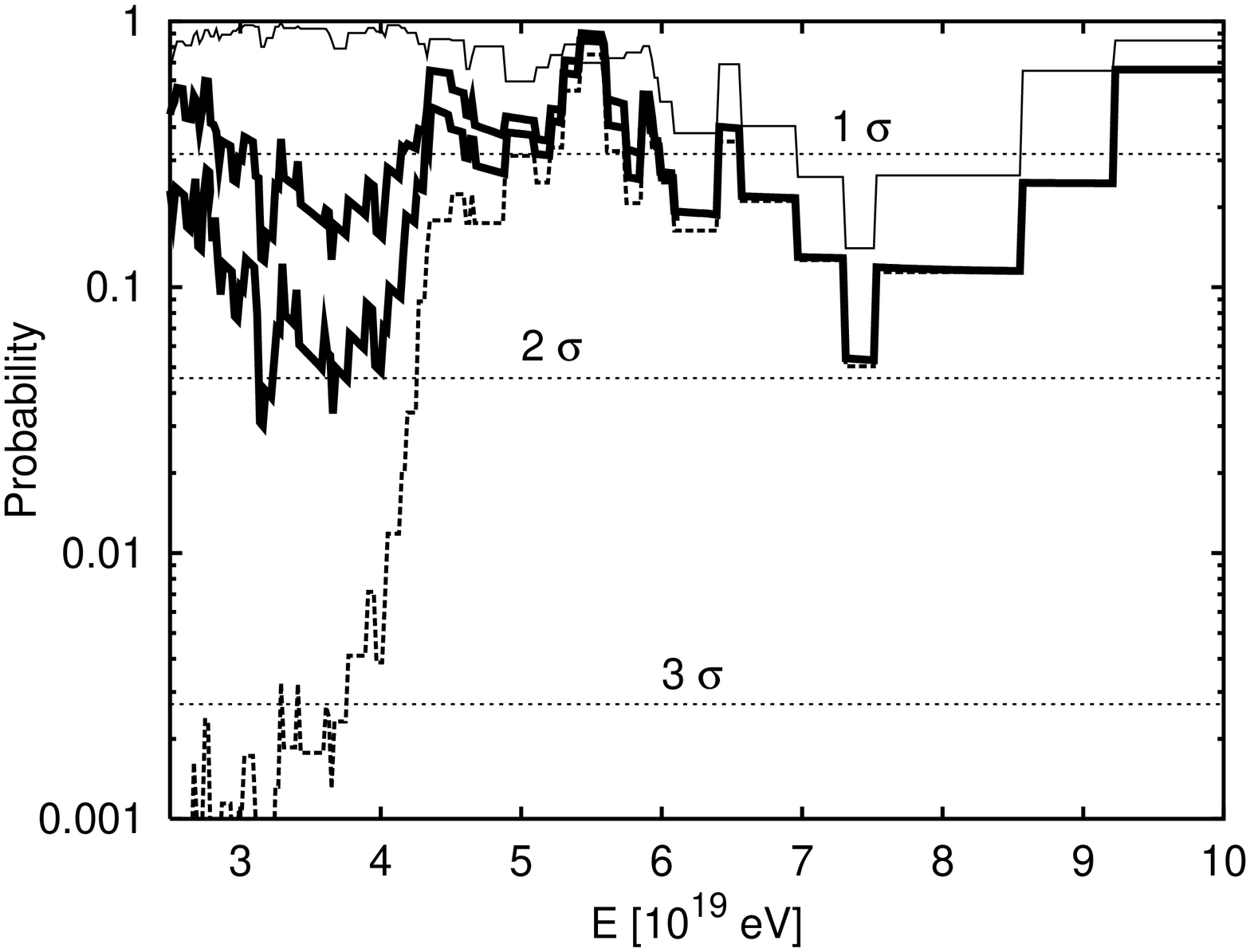}
\includegraphics[height=0.45\textwidth,clip=true,angle=270]{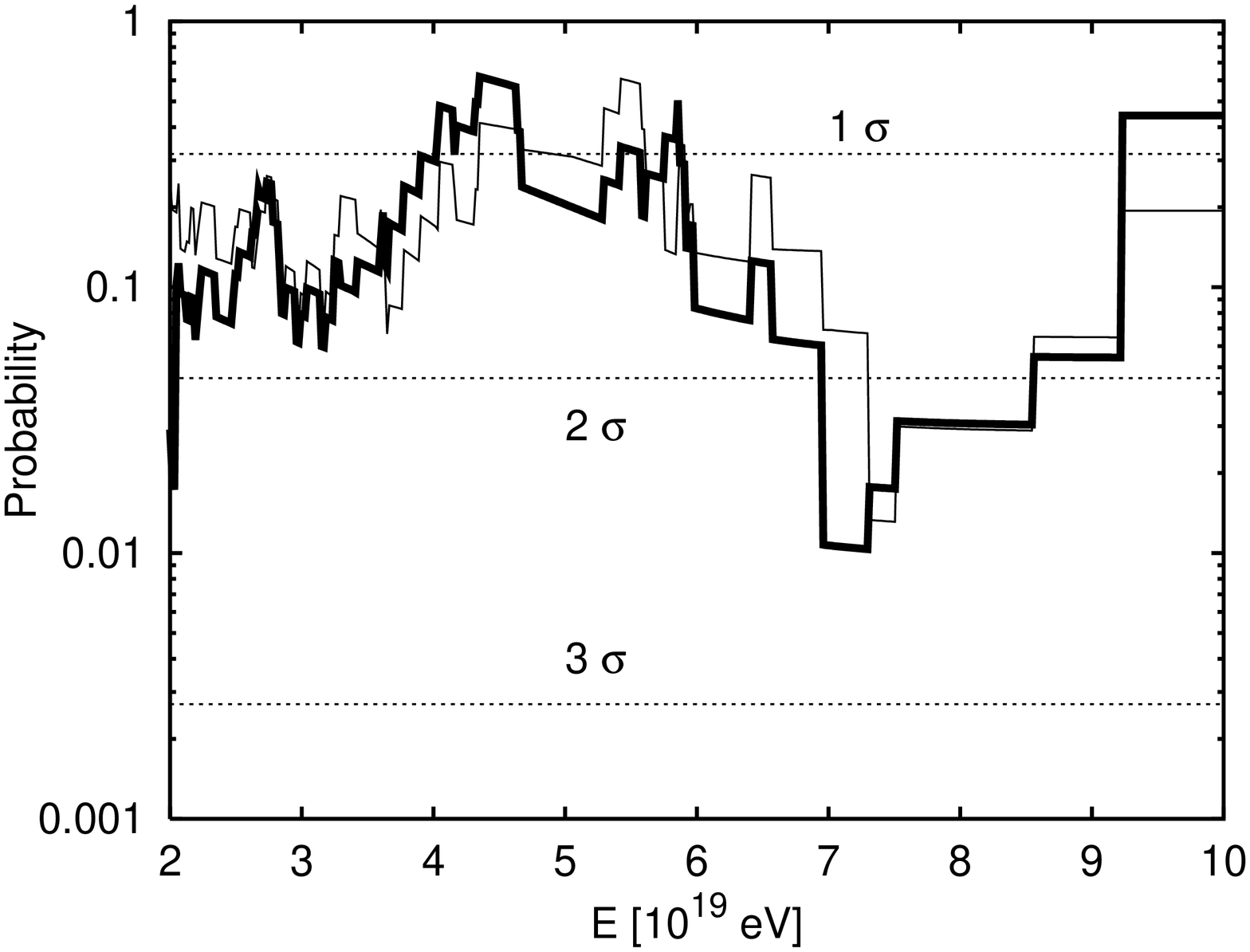}
\caption[...]{Left: Dependence of the probability on the energy cutoff in
  the SUGAR data for decaying SHDM.  
  Right: Two-dimensional KS test give results similar to one-dimensional test. 
\label{F04}}
\end{center}
\end{figure}

In Fig.~\ref{F04}a, we show the dependence of the probability on the
energy cutoff for $R_c=15$~kpc and $\theta_{\max}=55^\circ$. 
The two thick solid lines show $p$ for a combination of SHDM and
uniform sources according Eq.~(\ref{spec_norm}); the upper one
corresponds to an injection spectrum $1/E^{2.3}$, the lower one to 
an injection spectrum $1/E^{2.7}$. The behaviour of $p$
suggest that the minimum for $E_{\rm min}\sim 7\times 10^{19}$~eV is a
fluctuation similar to the maximum around  $E_{\rm min}\sim 5\times
10^{19}$~eV. In the range $E_{\rm min}\sim (3-4)\times 10^{19}$~eV,
the fluctuations adding an additional event are relatively
small. Therefore, we consider the probability in this range as more
reliable indicator for the consistency of SHDM with the SUGAR arrival
directions; we conclude that the SUGAR data have  the probability
$p=5-20\%$ to be consistent with the SHDM depending on the injection
spectrum of the extragalactic protons. 

The thin solid line shows how consistent the SUGAR data are with an
isotropic distribution. This distribution has also a minimum around
$E_{\rm min}\sim 7\times 10^{19}$~eV  where the events cluster around
two arrival directions. After including more low-energy data, the
SUGAR arrival directions are consistent with an isotropic distribution.   
Finally, the dashed line shows the consistency of the SUGAR data with the
assumption that all UHECR events above $E_{\rm min}$ are from SHDM. 
It is clear that only values of $E_{\rm min}$ above $E_{\rm min}\sim
4\times 10^{19}$~eV are compatible with the SUGAR data.  
Similar, the spectral shape of the flux in the SHDM model allows a
dominance of SHDM in the UHECR spectrum only above $E>6\times
10^{19}$~eV~\cite{FF}. 
In Fig.~\ref{F04}b we compare results from one- and two-dimensional
KS tests (of $\alpha$ and $\delta$) as function of the energy cutoff
and find that they give rather similar results.

\begin{figure}[ht]
\begin{center}
\includegraphics[height=0.45\textwidth,clip=true,angle=270]{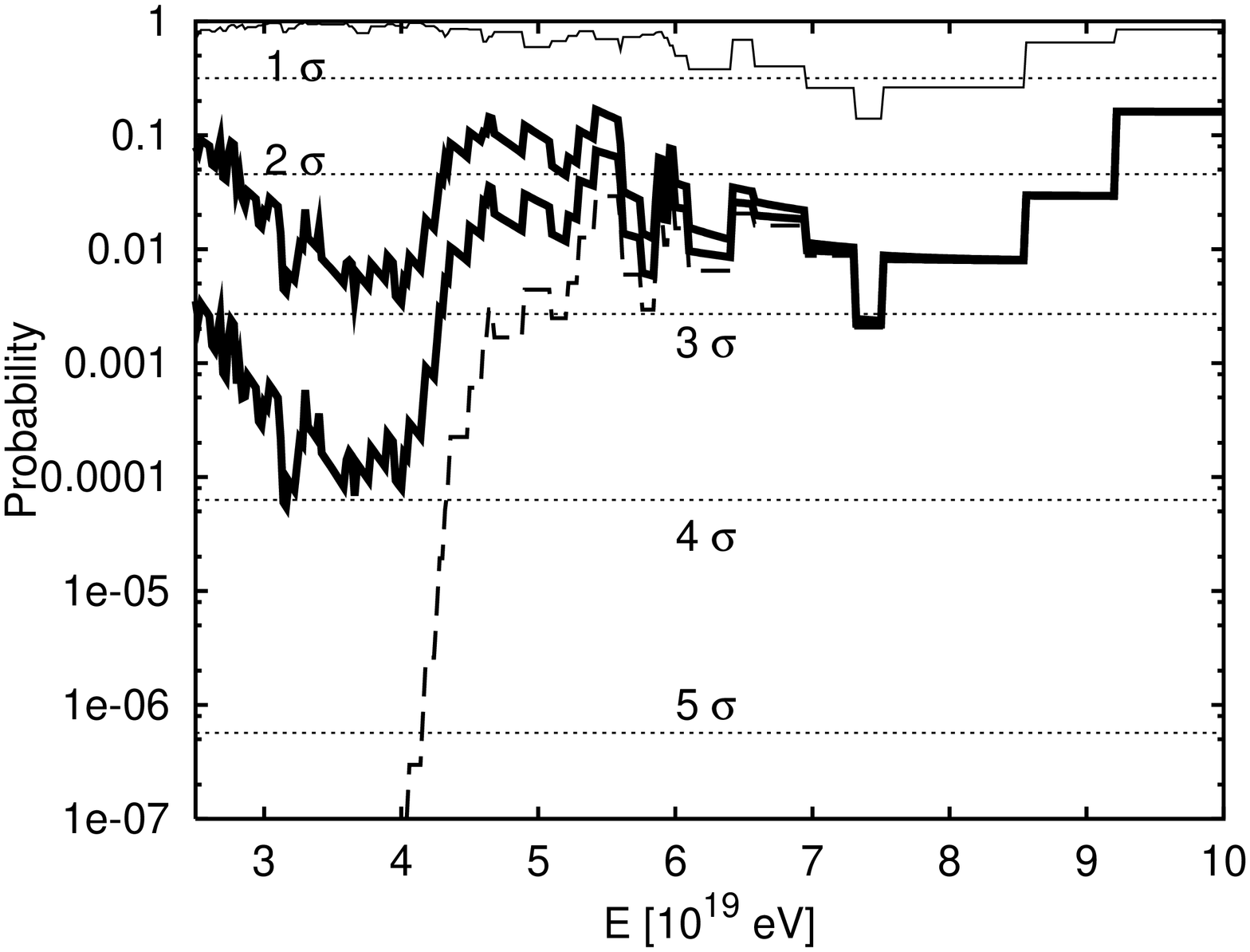}
\includegraphics[height=0.45\textwidth,clip=true,angle=270]{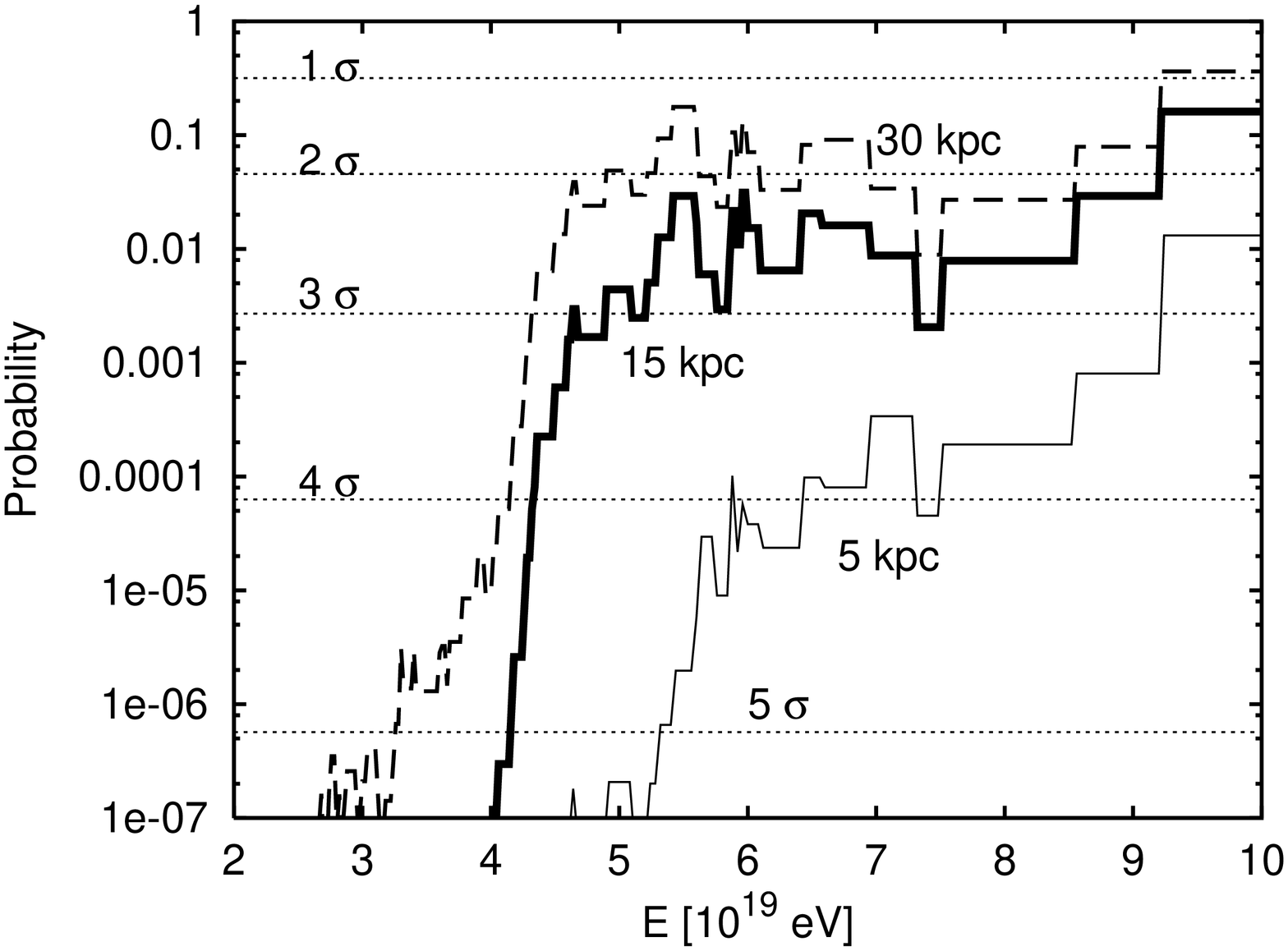}
\caption[...]{Dependence of the probability on the energy cutoff in
  the SUGAR data for annihilations of SHDM. 
  Left: for core radius $R_c=15$~kpc and different
  $\epsilon$ determining the SHDM contribution. 
  Right: for several core radii $R_c$; assumes that all events above
  $E$ are from SHDM.    
\label{F05}}
\end{center}
\end{figure}

Finally, we consider the model where not decays but annihilations of
SHDM particles produce the observed UHECRs~\cite{BDK}. In the original
version of this model it was suggested that the flux of the clumpy
component dominates over the one from the smooth SHDM profile by 3
orders of magnitude. On the other side, it was shown in a recent
paper~\cite{BDE_2003} that the contribution of the clumpy component
can be just a factor few larger than the one of the smooth component.
Moreover, the newest numerical calculations show that the contribution
of clumps is even subdominant and that it is very unlikely that a
nearby clump will outshine the Galactic center~\cite{private}.
Because of the arguments above, we assume that the clumpy part of the
SHDM gives a subdominant contribution to the 
UHECR flux. In the opposite case our results for the SHDM model with
annihilations will be less significant, depending on the relative
contribution of the two components.

Since the flux is now $\propto n^2_{\rm DM}$, the
anisotropy in this model is much stronger than for decaying SHDM.
This can be clearly seen in Fig.~\ref{F05}b where we show the
dependence of the probability of annihilating SHDM on the core radius
$R_c$ assuming that all events above $E$ are from SHDM. Even for core
radii as large as 30~kpc, annihilating SHDM is disfavoured by two
sigma. 
Figure~\ref{F05}a shows similar to Fig.~\ref{F04}a the dependence of
the probability on the energy cutoff for $R_c=15$~kpc and
$\theta_{\max}=55^\circ$. 
The two thick solid lines show $p$ for a combination of SHDM and
uniform sources according Eq.~(\ref{spec_norm}); the upper one
corresponds to an injection spectrum $1/E^{2.3}$, the lower one to 
an injection spectrum $1/E^{2.7}$. Depending on the injection spectrum
of extragalactic protons, annihilations of SHDM are disfavoured by the
SUGAR data between 3 and 4$\sigma$.

\section{Conclusions}

In this paper we have tested the consistency of the SHDM model with the 
SUGAR data.  In order to use the SUGAR data, we have compared its
energy spectrum to the one of AGASA and found that they are compatible
after rescaling down the SUGAR energies by 15\%. We have assumed that
the energy spectrum in the region $10^{19} ~{\rm eV} < E < 6\times
10^{19}$, i.e. between ankle and GZK cutoff, is dominated by protons
coming from uniformly distributed  extragalactic sources.
After fitting the relative contributions of SHDM decay products and
extragalactic protons to the AGASA data, we have performed
Kolmogorov-Smirnov tests of the SUGAR data.
As result we have found that SUGAR data are able to disfavour strongly
extreme case like annihilations of SHDM without clumps ($5\sigma$) or
decaying SHDM 
(99\% CL) assuming their contribution to the UHECR flux dominates down to
$E=4\times 10^{19}$~eV. The phenomenologically most interesting case,
decaying SHDM dominating the UHECR spectrum only above $E>6\times
10^{19}$~eV, is consistent with the SUGAR date with 5--20\%
probability. Thus the SUGAR data do not disfavour strongly this model
but they neither support it.    
A statistically significant test of this model can be done  
by the Pierre Auger Observatory.

\section*{Acknowledgements}

We would like to thank L.~Anchordoqui, R.~Dick, J.~Primack,
K.~Shinozaki, M.~Teshima, S.~Troitsky and S.~White for discussions
and comments. 
This work was supported by the Deutsche Forschungs\-ge\-meinschaft
(DFG) within the Emmy Noether program.

\vskip0.3cm
\noindent
{\em Note added:\/} 
After completing this work we have learnt about a preprint of 
H.~B.~Kim and P.~Tinyakov~\cite{kt} discussing also the
consequences of the SUGAR data for the SHDM hypothesis. These authors
come to similar conclusion as ours. We thank H.~B.~Kim  and P.~Tinyakov for
sending us the preprint before publication.



\begin{thebibliography}{99}



\bibitem{gzk} 
K.~Greisen,
Phys.\ Rev.\ Lett.\  {\bf 16}, 748 (1966).
G.~T.~Zatsepin and V.~A.~Kuzmin,
JETP Lett.\  {\bf 4}, 78 (1966)
[Pisma Zh.\ Eksp.\ Teor.\ Fiz.\  {\bf 4}, 114 (1966)].



\bibitem{agasa}
M.~Takeda {\it et al.},
Phys.\ Rev.\ Lett.\  {\bf 81}, 1163 (1998), 
[astro-ph/9807193];
N.~Hayashida {\it et al.},
Astrophys.\ J.\  {\bf 522}, 225 (1999)
[arXiv:astro-ph/0008102].
See also
{\sf http://www-akeno.icrr.u-tokyo.ac.jp/AGASA/}.


\bibitem{AGASAcluster_data}
M.~Takeda {\it et al.},
Astophys. J. {\bf 522} 225,
[arXiv:astro-ph/9902239];
Y.~Uchihori, M.~Nagano, M.~Takeda, M.~Teshima, J.~Lloyd-Evans and
A.~A.~Watson, 
Astropart.\ Phys.\  {\bf 13}, 151 (2000)
[arXiv:astro-ph/9908193];
\bibitem{AY_cluster_data}
P.~G.~Tinyakov and I.~I.~Tkachev,
JETP Lett.\  {\bf 74}, 1 (2001)
[Pisma Zh.\ Eksp.\ Teor.\ Fiz.\  {\bf 74}, 3 (2001)]
[arXiv:astro-ph/0102101].


\bibitem{Dubovsky2000}
S.~L.~Dubovsky, P.~G.~Tinyakov and I.~I.~Tkachev,
Phys.\ Rev.\ Lett.\  {\bf 85}, 1154 (2000)
[arXiv:astro-ph/0001317].

\bibitem{kst2003}
M.~Kachelrie{\ss}, D.~V.~Semikoz and M.~A.~Tortola,
hep-ph/0302161, to appear in Phys. Rev. {\bf D}.


\bibitem{Hires} 
D.~Kieda {\it et al.}, 
Proc. of the 26th ICRC, Salt Lake, 1999~, see also 
{\sf http://www.physics.utah.edu/Resrch.html}.
T.~Abu-Zayyad {\it et al.}  [High Resolution Fly's Eye Collaboration],
astro-ph/0208243.

\bibitem{corr_bllac}
P.~G.~Tinyakov and I.~I.~Tkachev,
JETP Lett. {\bf 74}, 445 (2001) 
[Pisma Zh.Eksp.Teor.Fiz. {\bf 74}, 499 (2001)]
[astro-ph/0102476];
%
see also 
P.~Tinyakov and I.~Tkachev,
astro-ph/0301336.


\bibitem{bkv97}
V.~Berezinsky, M.~Kachelrie{\ss}\ and A.~Vilenkin,
Phys.\ Rev.\ Lett.\  {\bf 79}, 4302 (1997)
[astro-ph/9708217].


\bibitem{kr97}
V.~A.~Kuzmin and V.~A.~Rubakov,
Phys.\ Atom.\ Nucl.\  {\bf 61}, 1028 (1998)
[Yad.\ Fiz.\  {\bf 61}, 1122 (1998)]
[astro-ph/9709187].


\bibitem{kt99}
V.~Kuzmin and I.~Tkachev,
Phys.\ Rev.\ D {\bf 59}, 123006 (1999)
[hep-ph/9809547];
D.~J.~Chung, E.~W.~Kolb and A.~Riotto,
Phys.\ Rev.\ D {\bf 60}, 063504 (1999)
[hep-ph/9809453].

\bibitem{dt98}
S.~L.~Dubovsky and P.~G.~Tinyakov,
JETP Lett.\  {\bf 68}, 107 (1998)
[hep-ph/9802382].

\bibitem{bbv98}
V.~Berezinsky, P.~Blasi and A.~Vilenkin,
Phys.\ Rev.\ D {\bf 58}, 103515 (1998).

\bibitem{photon_limit}
M.~Ave, J.~A.~Hinton, R.~A.~Vazquez, A.~A.~Watson and E.~Zas,
Phys.\ Rev.\ Lett.\  {\bf 85}, 2244 (2000)
[astro-ph/0007386];
K.~Shinozaki {\it et al.} [AGASA Collaboration],
Astrophys. J. {\bf 571}, 117 (2002).

\bibitem{an1}
A.~Benson, A.~W.~Wolfendale and A.~Smialkowski,
Astropart.\ Phys.\  {\bf 10}, 313 (1999).

\bibitem{an2}
L.~A.~Anchordoqui, C.~Hojvat, T.~P.~McCauley, T.~C.~Paul, S.~Reucroft,
J.~D.~Swain and A.~Widom, 
astro-ph/0305158.


\bibitem{BDK}
P.~Blasi, R.~Dick and E.~W.~Kolb,
Astropart.\ Phys.\  {\bf 18}, 57 (2002)
[astro-ph/0105232].


\bibitem{BGG}
 V.~Berezinsky, A.~Z.~Gazizov and S.~I.~Grigorieva,
astro-ph/0210095.

\bibitem{FF}
V.~Berezinsky and M.~Kachelrie\ss,
Phys.\ Rev.\ D {\bf 63}, 034007 (2001)
[hep-ph/0009053];
R.~Aloision, V.~Berezinsky and M.~Kachelrie\ss,
in preparation.


\bibitem{Winn:un}
M.~M.~Winn, J.~Ulrichs, L.~S.~Peak, C.~B.~Mccusker and L.~Horton,
J.\ Phys.\ G {\bf 12}, 653 (1986);
see also the complete catalogue of SUGAR data in
``Catalogue of highest energy cosmic rays No. 2'',  
ed.  WDC-C2 for Cosmic Rays (1986).


\bibitem{Winn:up}
M.~M.~Winn, J.~Ulrichs, L.~S.~Peak, C.~B.~Mccusker and L.~Horton,
J.\ Phys.\ G {\bf 12}, 675 (1986).

\bibitem{ave}
M.~Ave, J.~Knapp, J.~Lloyd-Evans, M.~Marchesini and A.~A.~Watson,
Astropart.\ Phys.\  {\bf 19}, 47 (2003)
[astro-ph/0112253].

\bibitem{muon}
A.~V.~Plyasheshnikov and F.~A.~Aharonian,
J.\ Phys.\ G {\bf 28}, 267 (2002)
[astro-ph/0107592].


\bibitem{QGS}
N.N.~Kalmykov, S.S.~Ostapchenko and A.I.~Pavlov,
Nucl. Phys. (Proc. Suppl.) {\bf 52B}, 17 (1997); 
N.N.~Kalmykov and S.S.~Ostapchenko, 
Preprint INP MSU 98-36/537, Moscow 1998;
N.N.~Kalmykov, S.S.~Ostapchenko and A.I.~Pavlov,
Izv. RAN Ser. Fiz. {\bf 58}, 21 (1994) 
(English translation in Bull.Russ.Acad.Sci (USA), 
Phys.Ser. {\bf v.58} 1966 (1994).)

\bibitem{linsley}
J.~Linsley,
Phys. Rev. Lett. {\bf 34}, 1530 (1975).


\bibitem{numrec}
W.~H.~Press, S.~A.Teukolsky, W.~T.~Vetterling and B.~P.~Flannery,
Numerical Recipes in Fortran, 
Cambridge University Press (1986, Cambridge).


\bibitem{NFW}
J.~F.~Navarro, C.~S.~Frenk and S.~D.~White,
Astrophys.\ J.\  {\bf 462}, 563 (1996)
[astro-ph/9508025].


\bibitem{code2} 
O.~E.~Kalashev, V.~A.~Kuzmin and D.~V.~Semikoz,
astro-ph/9911035;
Mod.\ Phys.\ Lett.\ A {\bf 16}, 2505 (2001)
[astro-ph/0006349].

\bibitem{milkyway}
A.~Klypin, H.~Zhao and R.~S.~Somerville,
astro-ph/0110390.

\bibitem{BDE_2003}
V.~Berezinsky, V.~Dokuchaev and Y.~Eroshenko,
astro-ph/0301551.

\bibitem{private}
F.~Stoehr, S.~D.M.~White, V.~Springel, G.~Tormen and N.~Yoshida,
astro-ph/0307026.

\bibitem{kt}
H.~B.~Kim and P.~Tinyakov,
astro-ph/0306413.


\end{thebibliography}
\end{document}